\documentclass[pra,twocolumn,preprintnumbers,amsmath,amssymb,
nofootinbib,floatfix]{revtex4}

\usepackage{graphicx,bm,amsmath}

\newcommand{\tr}[2]{\mathrm{Tr}_{#1}\left[#2\right]}

\makeatletter
\def\graphicscale{\twocolumn@sw{0.33}{0.4}}
\def\graphicthreescale{\twocolumn@sw{0.33}{0.4}}

\makeatother

\begin{document}

\title{Entanglement entropies in free fermion gases for arbitrary dimension}

\author{
Pasquale Calabrese, Mihail Mintchev and Ettore Vicari
}
\affiliation{Dipartimento di Fisica dell'Universit\`a di Pisa and INFN, 
Pisa, Italy}

\date{\today}

\begin{abstract}

We study the entanglement entropy of connected bipartitions in free
fermion gases of $N$ particles in arbitrary dimension $d$.  We show
that the von Neumann and R\'enyi entanglement entropies grow
asymptotically as $N^{(d-1)/d} \ln N$, with a prefactor that is
analytically computed using the Widom conjecture both for periodic and
open boundary conditions.  The logarithmic correction to the power-law
behavior is related to the area-law violation in lattice free
fermions.  These asymptotic large-$N$ behaviors are checked against
exact numerical calculations for $N$-particle systems.

\end{abstract}


\maketitle


The study of the entanglement properties of many body quantum systems
has attracted much attention in recent years, mainly in relation with
the critical behavior displayed at quantum phase transitions
\cite{AFOV-08,ECP-10,rev-cc}.  In particular, lots of studies have
been devoted to quantify the highly nontrivial connections between
different parts of an extended quantum system, by computing von
Neumann (vN) or R\'enyi entanglement entropies of the reduced density
matrix $\rho_A={\rm Tr}_B \rho$ of a subsystem $A$ with respect to its
complement $B$.

While a good understanding of the entanglement has been achieved in
one-dimensional (1D) systems , where a leading logarithmic behavior
has been established at conformal invariant quantum critical points
\cite{holzhey,vidalent,CC-04}, in higher dimensions the scaling
behavior of the bipartite entanglement entropy is generally more
complicated, without a definite general scenario for the universal
behavior at a quantum phase transition, despite of an enormous effort
on the problem.  Earlier investigations in the black hole physics
\cite{scr} proposed the general validity of the so-called area law
\cite{ECP-10}: for $d>1$ the entanglement entropy is asymptotically
proportional to the surface of the area separating the two subsystems
$A$ and $B$.  The area law has been generally proven for {\it gapped}
systems~\cite{wolf}, independently of the statistics of the
microscopical constituents.  This result is corroborated by exact and
numerical calculations for systems with topological order~\cite{top},
at 2D quantum Lifshitz points~\cite{cftw}, and some explicit field
theoretical computations \cite{CH}.  The situation is less clear for
critical systems.  On the one hand, the area law holds in bosonic
systems~\cite{boso,BCS-06,ECP-10}.  Quantum critical points described
by a Landau-Ginzburg action satisfy the area law \cite{mfs-09}, which
is also supported by holographic calculations by means of AdS/CFT
\cite{holo} with subleading logarithmic corrections.  Even in some
disordered systems the area law is confirmed by analytical and
numerical calculations and the presence of universal subleading term
has been shown \cite{dis}.  On the other hand, the simplest condensed
matter models, i.e. free fermions on a lattice, show multiplicative
logarithmic corrections to the area law
\cite{Wolf-06,GK-06,BCS-06,LDYRH-06,FZ-07,HLS-09,DBYH-08,Swindle-10,dsy-11},
i.e.  for a large subsystem $A$ of linear size $\ell$ in an infinite
$d$-dimensional lattice the entanglement entropies scale like
\begin{equation}
S^{(\alpha)}(A) \sim \ell^{d-1} \ln \ell\,,
\label{sadl}
\end{equation}
where $S^{(\alpha)}(A)$ are the R\'enyi entropies defined as
\begin{equation}
S^{(\alpha)}(A) = \frac{1}{1-\alpha} \ln {\rm Tr}\rho_A^\alpha\,.
\label{saldef}
\end{equation}
For $\alpha\to 1$, one recovers the vN definition $S(A) \equiv
S^{(1)}(A) \equiv -\tr{}{\rho_A\ln\rho_A}$.  It has been argued that
these logarithmic corrections should also appear for interacting
fermions with a finite Fermi
surface~\cite{Wolf-06,GK-06,LDYRH-06,FZ-07,HLS-09,DBYH-08,Swindle-10,dsy-11}.

In this paper we investigate the entanglement properties of free
fermion gases of $N$ particles in arbitrary dimensions, focussing on
the large-$N$ asymptotic behavior of the entanglement entropies of
connected spatial bipartitions. Analogously to lattice models, an
interesting issue concerns the existence of multiplicative logarithmic
corrections to the large-$N$ power-law behavior corresponding to the
area law. A related question is whether these logarithmic corrections
are due to the presence of corners in the area separating $A$ and $B$,
or they are also present for a smooth surface.  

In Ref. \cite{CMV-11} we developed a systematic framework to calculate
the entanglement entropies of free fermion gases for a finite number
of particles $N$, based on the overlap matrix of the lowest $N$ states.
This method has been already applied to 1D systems
\cite{CMV-11,CMV-11a,CMV-11b}, but it allows also computations in higher
dimensions.  We consider systems of $N$ particles in a finite volume
of arbitrary dimension $d$, and focus on the large-$N$ scaling
behavior of the spatial entanglement entropies of connected
bipartitions, finding a general $N^{(d-1)/d}\ln N$ behavior for large
$N$.  This multiplicative logarithmic correction to the large-$N$
power law is related to the logarithmic correction to the area law in
lattice models.  Combined with simple scaling arguments, this result
allows us to conclude that multiplicative logarithmic corrections to
the area law are general features of free-fermion systems both on the
lattice and in the continuum.

{\it The method}.  The ground-state wave function of a gas of $N$
noninteracting spinless fermions is $\Psi({\bf x}_1,...,{\bf x}_N) =
{\rm det} [\psi_i({\bf x}_j)]/\sqrt{N!}$, where $\psi_i$ are the
normalized wave functions of the one-particle problem with lowest
energies.  Explicit expressions of the one-particle eigensolutions are
products of eigenfunctions of corresponding 1D Sch\"rodinger problems,
i.e.  $\psi_{n_1,n_2,...,n_d}({\bf x}) = \prod_{i=1}^d \phi_{n_i}(x_i)
$, where the $n_i$ label the eigenfunctions along the $d$ directions
which are solution of $- \partial_x^2 \phi_{n}(x) = e_{n}\phi_n(x)$,
so that the energy of the $d$-dimensional problem is $
E_{n_1,n_2,...,n_d}= \sum_{i=1}^d e_{n_i}$.  Note that,
although the spatial dependence of the one-particle eigenfunctions is
essentially decoupled along the various directions, fermion gases in
different dimensions present notable differences due to the nontrivial
filling of the lowest $N$ states to obtain the ground state of the
$N$-particle system.

In the following we consider hypercubic systems of
linear size $L$ with periodic boundary conditions (PBC)
and Dirichelet (open) boundary conditions (OBC). 
The corresponding normalized 1D eigensolutions are
$\phi^{\rm PBC}_k(x)=L^{-1/2} e^{2\pi i k x/L}$ for PBC and $\phi^{\rm
OBC}_k(x)=(2/L)^{1/2}\sin k \pi x/L$ for OBC, with $k\in {\mathbb Z}$
and $x\in[0,L]$.
 
The spatial entanglement entropies of connected bipartitions can be
computed using the method developed in Refs.~\cite{CMV-11,CMV-11a} (to
which we refer for details).  For this purpose, we consider the
$N\times N$ {\em overlap} matrix ${\mathbb A}$ with elements
\begin{equation}
{\mathbb A}_{nm} =  \int_A d^d z\, \psi_n^*({\bf z}) \psi_m({\bf z}),
\qquad n,m=1,...,N,
\label{aiodef}
\end{equation}
where the integration is over the spatial region $A$, and involves the
lowest $N$ energy levels.  Then, the R\'enyi entanglement entropies
are derived from the formula
\begin{equation}
S^{(\alpha)}(A) = \sum_{n=1}^N e_\alpha(a_n),
\label{snx2n}
\end{equation}
where $a_n$ are the eigenvalues of ${\mathbb A}$, and
\begin{equation}
e_\alpha(\lambda) = {1\over 1-\alpha} \ln \left[{\lambda}^\alpha
+\left({1-\lambda}\right)^\alpha\right].
\label{enx}
\end{equation}
The vN entropy is obtained by the limit $\alpha\to 1$.  The knowledge
of the $S^{(\alpha)}(A)$ for different $\alpha$ characterizes the full
spectrum of non-zero eigenvalues of $\rho_A$ \cite{cl-08}, and gives
more information about the entanglement than the single vN entropy.
Furthermore, numerical stochastic methods such as classical
\cite{cg-08} and quantum Monte Carlo \cite{melko} can measure only
R\'enyi entropies with integer $\alpha$.

For systems with both PBC and OBC, the form of the overlap matrix is
analogous to that of the two-point correlation of free spinless
fermions on the lattice.  Indeed, taking as simpler example the PBC
case, the two-point function is
\begin{equation}
G_{\rm lat}({\bf x},{\bf y}) = \int_{\Gamma(\mu)} {d^d k \over (2\pi)^d} \;
e^{i {\bf k}  \cdot  ({\bf x}-{\bf y}) }
\label{twoplat}
\end{equation}
where $\Gamma(\mu)$ is the volume limited by the Fermi surface
$\partial\Gamma(\mu)$, and ${\bf x},{\bf y}$ are the positions of the
sites of the lattice in unit of the lattice spacing, thus ${\bf x}\in
{\mathbb Z}^d$.  On the other hand, after inserting the one-particle
eigensolutions, the overlap matrix (\ref{aiodef}) reads
\begin{equation}
{\mathbb A}_{nm} = \int_{A}  d^d x  \;
e^{i2\pi ({\bf k}_m-{\bf k}_n) \cdot  {\bf x} }\;.
\label{anmpb}
\end{equation}
Eqs.~(\ref{twoplat}) and (\ref{anmpb}) look very similar and present a
clear correspondence: in the lattice two-point function the integer
arguments are related to the sites and the integration over the
momenta is within the Fermi surface, while in the overlap matrix the
indices are related to the states and the integration is within the
spatial region $A$. Thus $N$ plays the role of the volume of the
subsystem in the lattice model.  This correspondence has been already
exploited to derive rigorous asymptotic large-$N$ results for the
bipartite entanglement entropy in 1D fermion
gases~\cite{CMV-11,CMV-11a}.

\begin{figure}[tbp]
\includegraphics*[scale=\graphicscale]{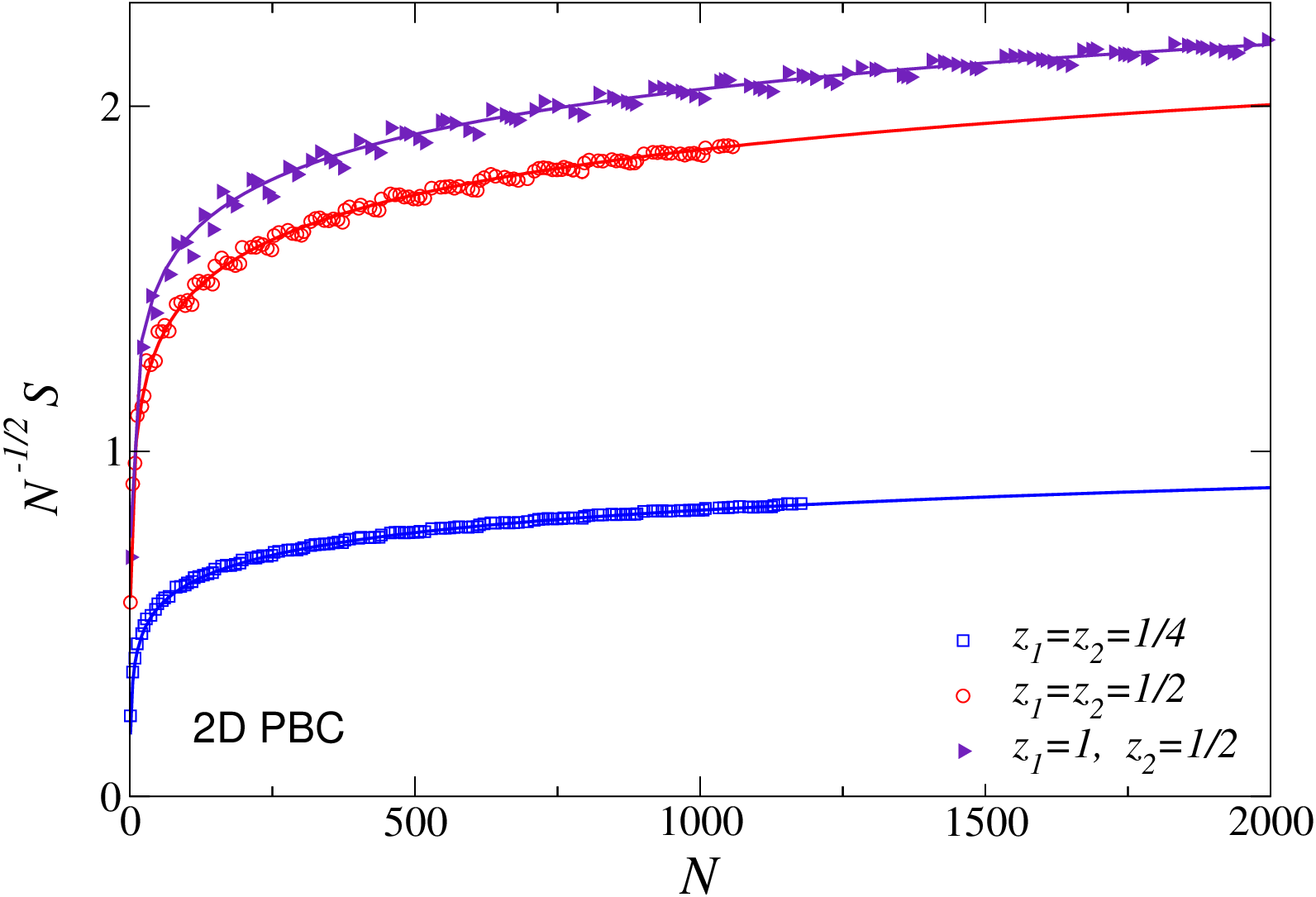}
\includegraphics*[scale=\graphicscale]{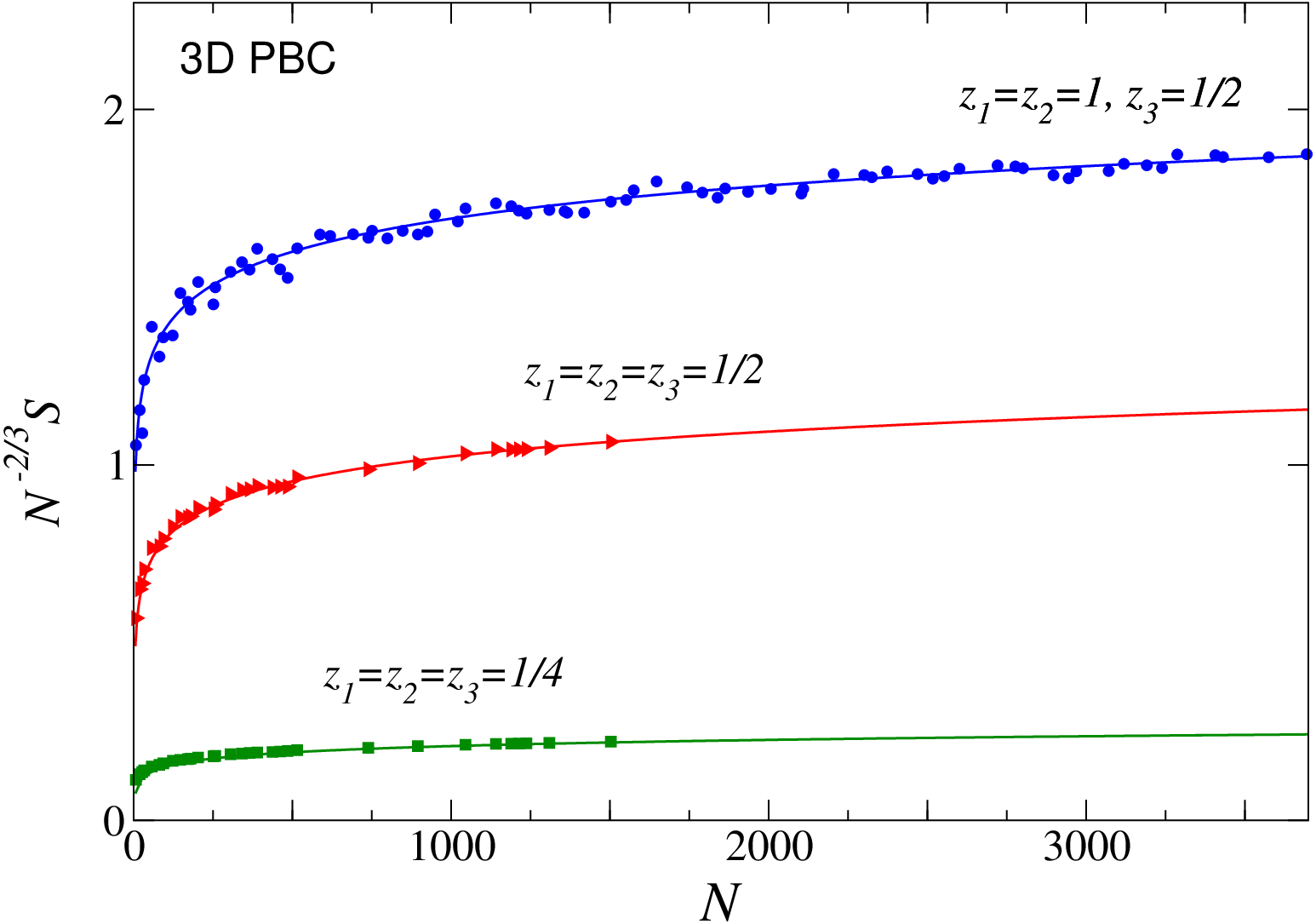}
\caption{(Color online) The bipartite vN entanglement entropy $S$ for
space regions $A=[0,z_1L]\times [0,z_2L]\dots \times [0,z_d L]$ 
in  2D (top) and 3D (bottom) systems  with PBC.  
We report $S/N^{1-1/d}$ vs $N$. 
The full lines show the asymptotic behavior
$S= c \ln N + b$ where $c$ is the geometry-dependent constant in
Eqs.~(\ref{resca2d1}) and (\ref{resca2d2}), while the constants $b$
are fitted from the data for the largest values of $N$.  }
\label{s1p2d}
\end{figure}

{\it Systems with periodic boundary conditions}.  To begin with we
consider systems with PBC.  We compute the entanglement entropies for
regions $A=[0,z_1L]\times [0,z_2L]\times ...\times [0,z_dL]$.  If all
$z_i<1$, $A$ is an hypercube with corners, but if all $z_i=1$ except
one (let us say $z_1<1$), then, because of PBC, the subsystem $A$ is
$d$ dimensional strip.  In order to calculate the entanglement
entropies, we construct the overlap matrix ${\mathbb A}$ from the
one-particle wave functions, calculate numerically its eigenvalues and
then use Eq.~(\ref{snx2n}) for $S^{(\alpha)}$.  Practically, the data
are obtained for increasing values of the energy $E$, and $N$ is
extracted by counting the states with $E_i\le E$.  The results in
$d=2$ and $d=3$ are reported in Fig.~\ref{s1p2d} for some values of
$z_i$ and for $N$ up to $O(10^3)$.  The plot of $S/N^{1-1/d}$ vs $N$
clearly confirms the presence of the multiplicative logarithmic corrections.

For the lattice free fermion model, the coefficient of the leading
large-$\ell$ behavior in Eq. (\ref{sadl}) has been evaluated by
assuming the validity of the Widom
conjecture~\cite{Widom-81,HLS-09}.  In the thermodynamic limit, the
bipartite entanglement entropy of the region $A$ can be computed from
the lattice two-point function (\ref{twoplat}) restricted to $A$.  In
the case of a connected space region $A$ of size $\ell$, the R\'enyi
entanglement entropies behave as~\cite{GK-06,HLS-09}
\begin{eqnarray}
{\cal S}^{(\alpha)}(A)&=& {1+\alpha^{-1}\over 2} c(\mu) \ell^{d-1} 
\ln \ell + o(\ell^{d-1}\ln \ell),\;\;\label{lpbc}\\
c(\mu) &=& {1\over 12 (2\pi)^{d-1}} \int_{\partial \Omega} dS_x
\int_{\partial\Gamma(\mu)} \hspace{-4mm} dS_k  |{\bf n}_k\cdot {\bf n}_x|,\;\;
\label{cmu}
\end{eqnarray}
where $\mu$ is the chemical potential, $\Omega$ is the real-space
region $A$ rescaled by $\ell$ so that its volume equals one.  ${\bf
n}_x$ and ${\bf n}_k$ denote the normal vectors on the spatial surface
$\partial\Omega$ and the Fermi surface $\partial\Gamma(\mu)$,
respectively.  The validity of these predictions has been supported by
numerical results~\cite{BCS-06,LDYRH-06}.

\begin{figure}[tbp]
\includegraphics*[scale=\graphicscale]{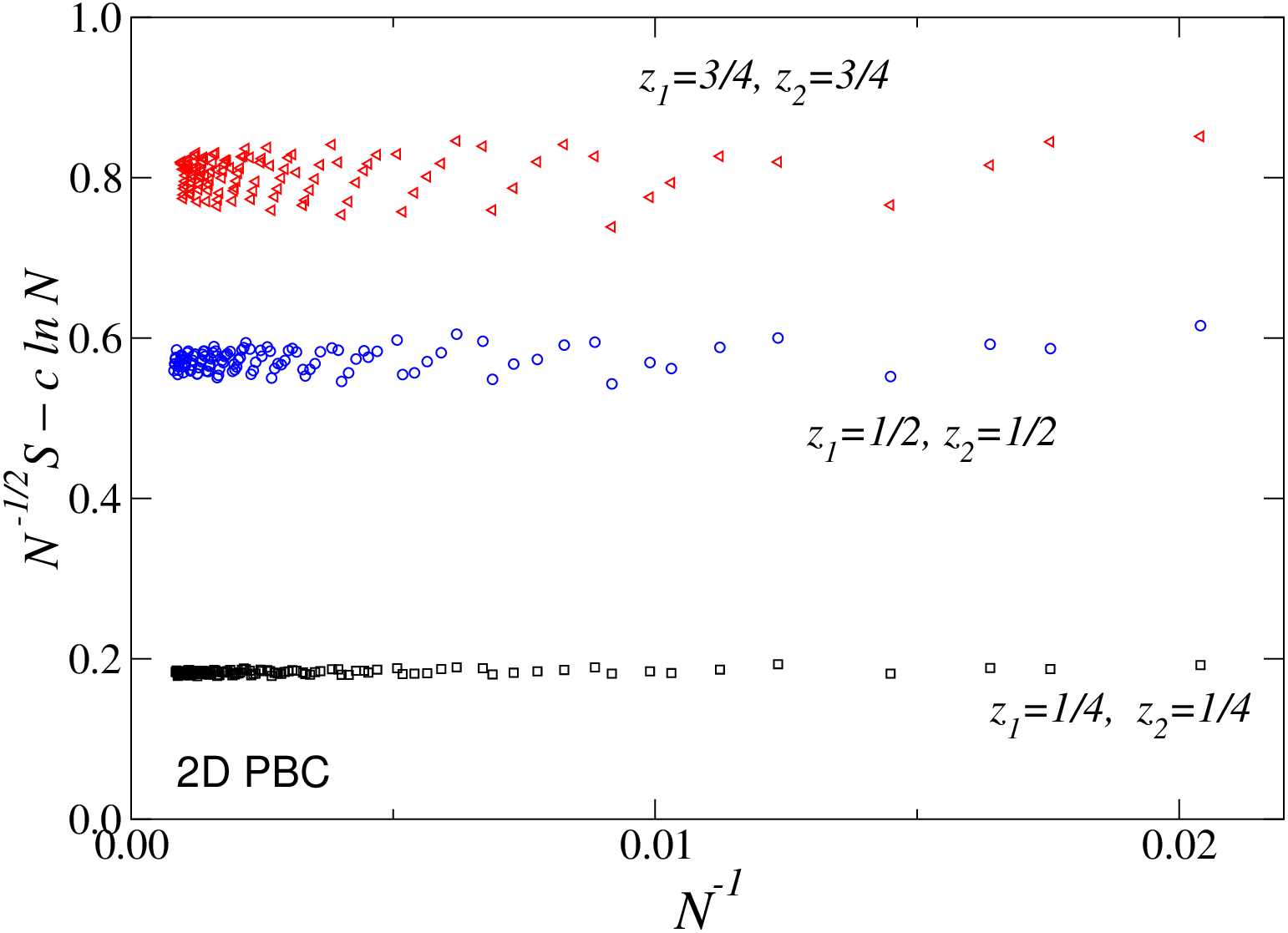}
\includegraphics*[scale=\graphicscale]{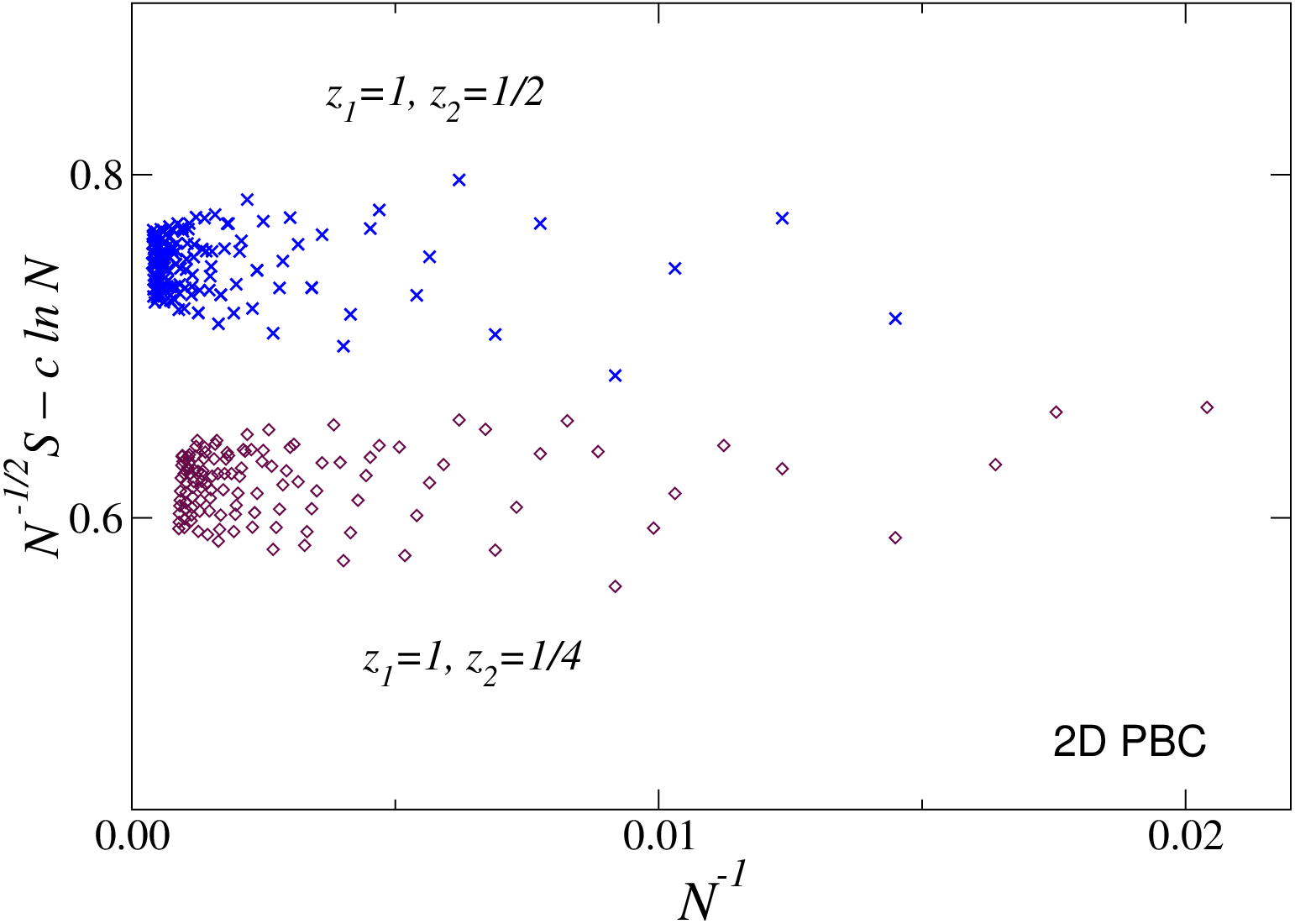}
\caption{(Color online) The bipartite vN entanglement entropy of 
space regions $A=[0,z_1L]\times [0,z_2L]$ in a 2D
system with PBC.  We plot the
quantity $S/N^{1/2} - c \ln N$ versus $N^{-1}$ for square regions
(top) and stripes (bottom). The coefficients $c$ are given by
Eqs.~(\ref{resca2d1}) and (\ref{resca2d2}) respectively.  }
\label{ee2dpb}
\end{figure} 

Calculations based on the Widom conjecture allow us to derive an
analogous result for the large-$N$ asymptotic behavior of the
entanglement entropy in fermion gases defined in the continuum, as
already suggested by the above-mentioned correspondence between the
overlap matrix ${\mathbb A}$ and the lattice two-point function. For a
$d$-dimensional cubic volume of size $L$ with PBC, containing $N$
fermions, the corresponding Fermi surface is a $d$-dimensional sphere
of radius $k_F$, which is related to the particle number $N$ by
\begin{equation}
V_\Gamma = v_d k_F^d = N,\qquad v_d=
{\pi^{d/2}\over \Gamma(d/2+1)} .
\label{vgamma}
\end{equation}
We want to compute the vN entanglement entropy of a space region $A$,
which may be given by the volume $[0,z_1L]\times [0,z_2L]\times
...\times [0,z_dL]$ with $z_i\le 1$.  Using the Widom conjecture as in
Ref.~\cite{GK-06}, we obtain the asymptotic large-$N$ behavior
\begin{eqnarray}
&&S(A) = c  N^{(d-1)/d} \ln N + o(N^{(d-1)/d}\ln N),
\quad\label{npbc}\\
&&c = {1\over 12d v_d^{(d-1)/d}} \int_{\partial \gamma} dS_\kappa
\int_{\partial a} dS_z  | {\bf n}_z\cdot {\bf n}_\kappa|,
\label{cxi}
\end{eqnarray}
where $\partial a$ is the surface of the rescaled volume
$[0,z_1]\times [0,z_2]\times ...\times [0,z_d]$, and $\partial \gamma$
is the Fermi surface rescaled to have a unit radius.  The integral
(\ref{cxi}) gives
\begin{equation}
c=\frac{ \Gamma[d/2]}{6 \Gamma \left[(d+1)/{2} \right](\Gamma [{d/2}+1])^{1/d}}
\sum_{j=1}^d \prod_{i\neq j} z_i\,,
\label{genc}
\end{equation}
and specifically
\begin{equation}
c=\left\{\begin{array}{lc}
\displaystyle \frac13\,,& d=1,\\
\displaystyle \frac1{3\sqrt{\pi}} (z_1+z_2),&d=2,\\
\displaystyle \frac16\left(\frac\pi{6}\right)^{1/3}(z_1z_2+z_1z_3+z_2z_3),&d=3.\\
\end{array}\right.
\label{resca2d1}
\end{equation}
Using the Widom conjecture, one can also derive analogous formulas for
generic shapes of the connected space region $A$.  Note that the
factor $2\sum_{j=1}^d \prod_{i\neq j} z_i$ is just the area of the
surface separating $A$ and $B$ in units of $L=1$.  This could lead to
the erroneous conclusion that the area law is satisfied in these
systems, but Eq. (\ref{npbc}) is just an asymptotic expansion in $N$
and the size-dependent multiplicative logarithmic correction to the
area law is a subleading term of the form $N^{1-1/d}$ times the
logarithm of the area (see the conclusions for a discussion of this
issue).

Note that Eq.~(\ref{genc}) is not valid when one $z_i$ is one,
because the terms at the edge do not contribute to the area,
i.e. the approach to the asymptotic large-$N$ behavior (\ref{npbc}) is
not uniform as a function of $z_i$.  In particular, when all $z_i=1$
except one (let us say $z_1$), the area separating $A$ and $B$ is just
$2$ (in units of $L$) and we have
\begin{equation}
c_S=\frac{ \Gamma \left(\frac{d}{2}\right)}
{6 \Gamma \left(\frac{d+1}{2} \right)\Gamma \left(\frac{d}{2}+1\right)^{1/d}}\,,
\label{resca2d2}
\end{equation}
independently of $z_1$.  In this case, the subsystem $A$ is a
$d$-dimensional strip without corners (because of PBC).  Note that the
approach to the asymptotic large-$N$ behavior (\ref{npbc}) is not
uniform as a function of $z_1$ as well, because for $z_1=0$ or $1$ the
entanglement entropy trivially vanishes.

The Widom conjecture does not provide information on the corrections
to the leading behavior.  As we shall see, the analysis of finite-$N$
computations with increasing $N$ indicates that they are
$O(N^{1-1/d})$, i.e.
\begin{eqnarray}
S(A) =  N^{(d-1)/d} \left[ c \ln N + O(1)\right],
\label{npbcc}
\end{eqnarray}
with $c$ given by Eq.~(\ref{cxi}). Note that setting $d=1$ into this
formula, we recover the rigorous 1D results for the
asymptotic entanglement entropy~\cite{CMV-11,CMV-11a}.  However, in 1D
the Fisher-Hartwig conjecture not only gives the asymptotic behavior
as above, but also the $O(1)$ term and the precise form of all
subleading corrections, both for particle systems
\cite{CMV-11,CMV-11a} and lattice models \cite{JK-04,CCEN-10}.

\begin{figure}[tbp]
\includegraphics*[scale=\graphicscale]{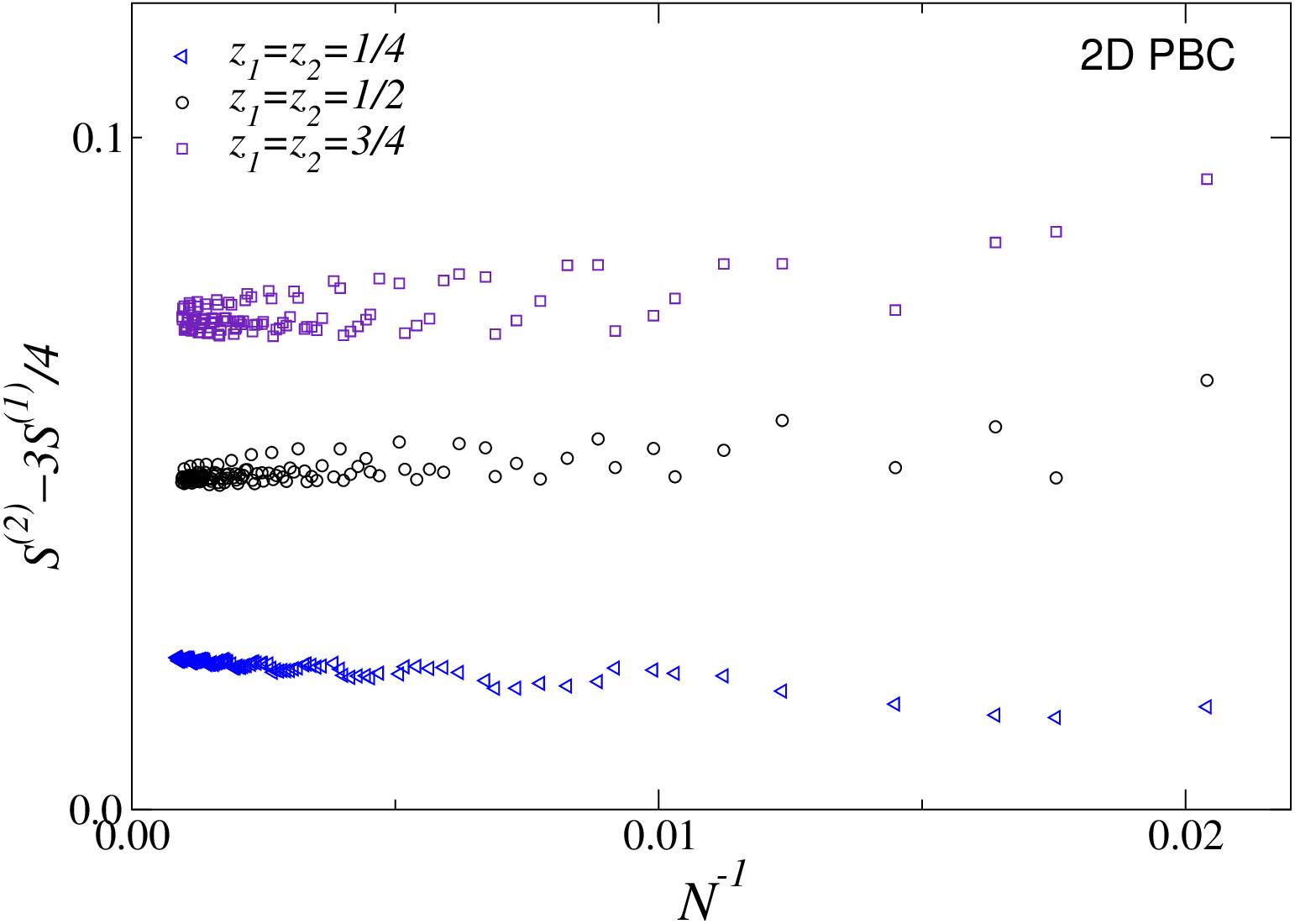}
\caption{(Color online) We plot $N^{-1/2}(S^{(2)}-3 S/4)$ built
using the $\alpha=2$ R\'enyi and vN entropies of spatial regions
$A=[0,z_1L]\times [0,z_2 L]$ for some values of $z_1,z_2$, versus
$N^{-1}$.  }
\label{2dpbs2os1}
\end{figure}

\begin{figure}[t]
\includegraphics*[scale=\graphicscale]{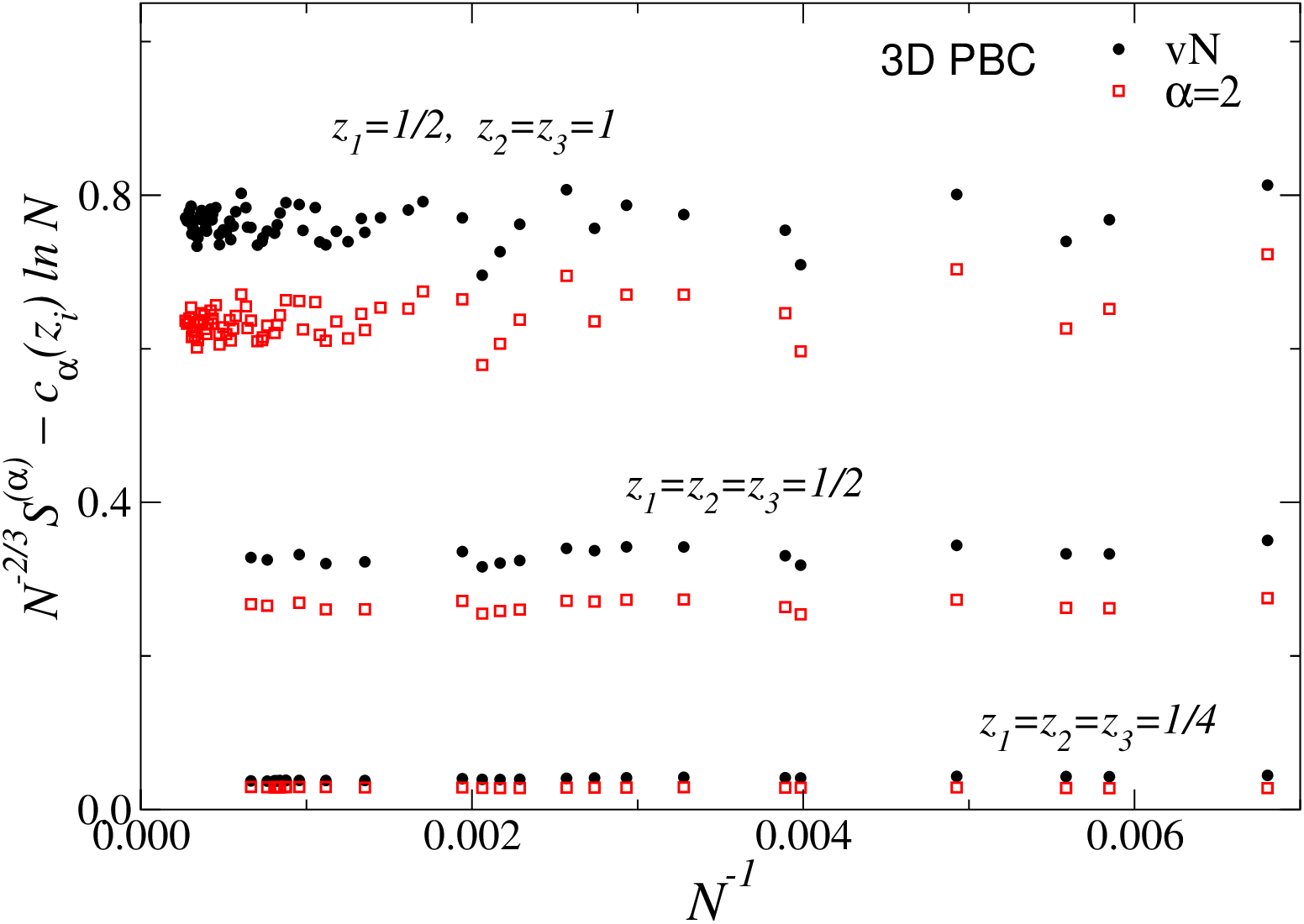}
\caption{(Color online)
The bipartite vN and R\'enyi entanglement entropy of 
space regions $A=[0,z_1L]\times [0,z_2L]\times [0,z_3L]$ in a 3D system with PBC.  
We plot the quantity $S/N^{2/3} - c_{\alpha} \ln N$ versus $N^{-1}$ for 
different values of $z_i$. 
}
\label{3dplot}
\end{figure}

The asymptotic behaviors predicted by Eqs.~(\ref{npbc}-\ref{cxi}) are
definitely supported by calculations of the bipartite vN entanglement
entropy, as shown by Fig.~\ref{s1p2d}.  To be more quantitative, in
Fig.~\ref{ee2dpb} we report for the difference
\begin{equation}
N^{-1/2}S(A)  - c \ln N
\label{diffdef}
\end{equation}
for 2D systems, with $c$ given by Eq.~(\ref{cxi}).  This difference
appears limited, although presenting sizable oscillations which
enlarge with increasing $z_1$ and $z_2$.  Its large-$N$ limit provides
information on the subleading behavior of $N^{-1/2}S$, which appears
to be $O(1)$, supporting Eq.~(\ref{npbcc}).

The amplitude of the oscillations modulating the difference
(\ref{diffdef}) appears to decrease. However  it is not clear how they
decrease and whether the amplitude of the oscillations get suppressed
in the large-$N$ limit.  Oscillating corrections to the leading
behavior have been observed also for 1D systems, but in the case of
PBC only for R\'enyi entropies with $\alpha\neq1$
\cite{CMV-11a,CCEN-10}.  Their origin can be traced back to the
presence of relevant operators localized at the borders of the region
$A$ \cite{CC-10}.  However, the ones observed in $d>1$ are
present also for $\alpha=1$ and their structure (periods, amplitudes
etc.) appear very different from the 1D ones.  For these reasons both
the general structure and their physical origin remains unclear
and this issue would deserve further investigations.

For the R\'enyi entropies, the application of the Widom conjecture 
leads to
\begin{equation}
S^{(\alpha)} = {1+\alpha^{-1}\over 2} S(A) + o(N^{(d-1)/d}\ln N)\,,
\label{rae}
\end{equation}
analogously to lattice models~\cite{HLS-09}.  This is supported by the data shown in
Fig.~\ref{2dpbs2os1}, where we plot the difference $N^{-1/2}(S^{(2)}-3
S/4)$ between the $\alpha=2$ R\'enyi and vN entropies of spatial
regions $A=[0,z_1L]\times [0,z_2 L]$ for some values of $z_1,z_2$.
The data appear to approach a constant but with large oscillations
that make difficult to extract the rate of convergence.

In Fig. \ref{3dplot} we report analogous results in $d=3$.  Again, the
subtracted quantity $N^{-2/3} S^{(\alpha)}-c_\alpha \ln N$ turns out
to be limited, indicating that it is $O(1)$ 
(in agreement with Eq. (\ref{npbcc})),  with sizable oscillating
corrections whose qualitative features are again unclear.

\begin{figure}[tb]
\includegraphics*[scale=\graphicscale]{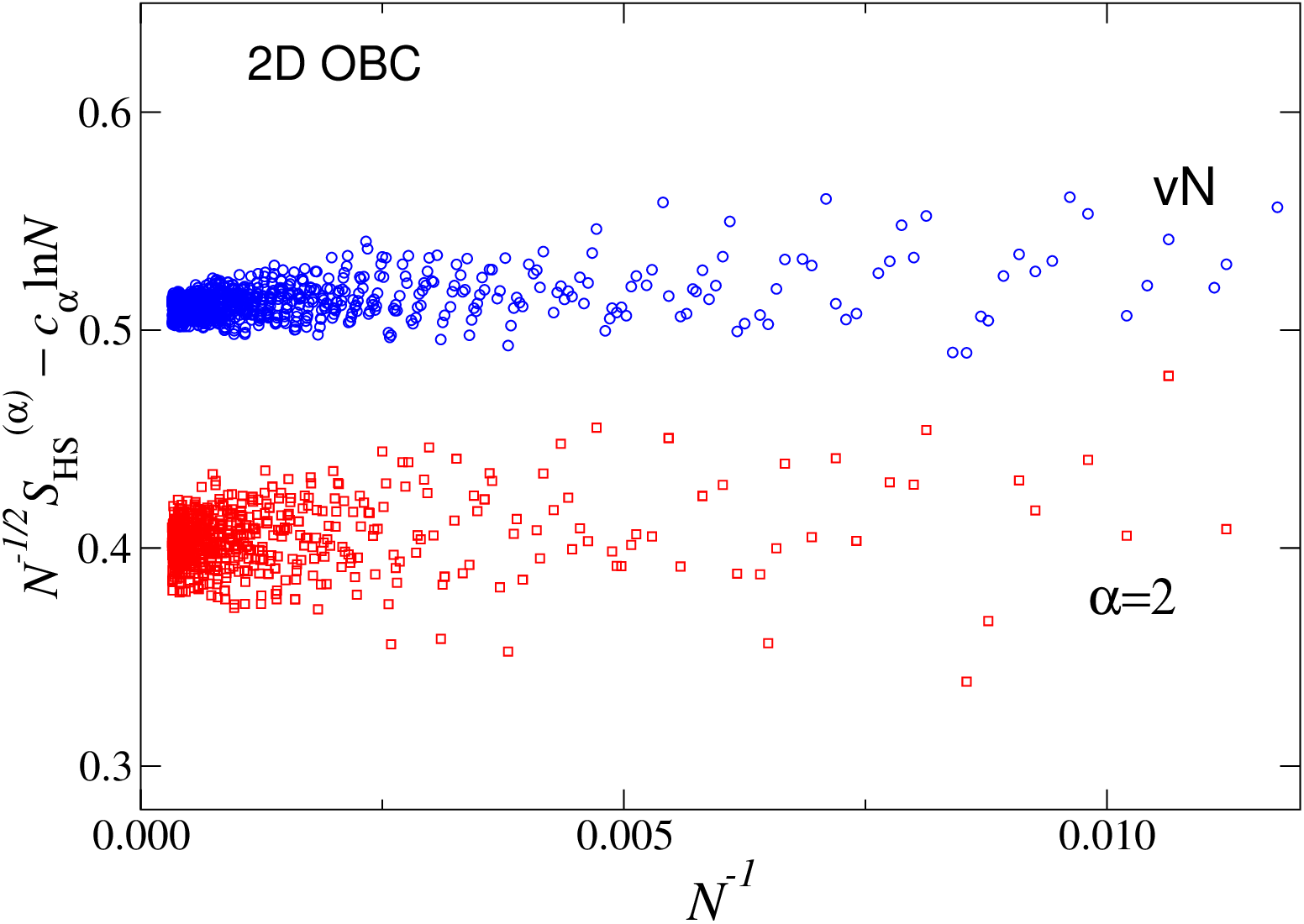}
\caption{(Color online) The half-space vN and $\alpha=2$ R\'enyi
entanglement entropies for  a
 system with OBC, up to $N\approx
3000$.  We plot the  $N^{-1/2}S_{\rm HS}^{(\alpha)} -
c_{\alpha} \ln N$ with $c_\alpha$ 
in Eq.~(\ref{uobca}).  }
\label{ee2dobhs}
\end{figure}

\begin{figure}[t]
\includegraphics*[scale=\graphicscale]{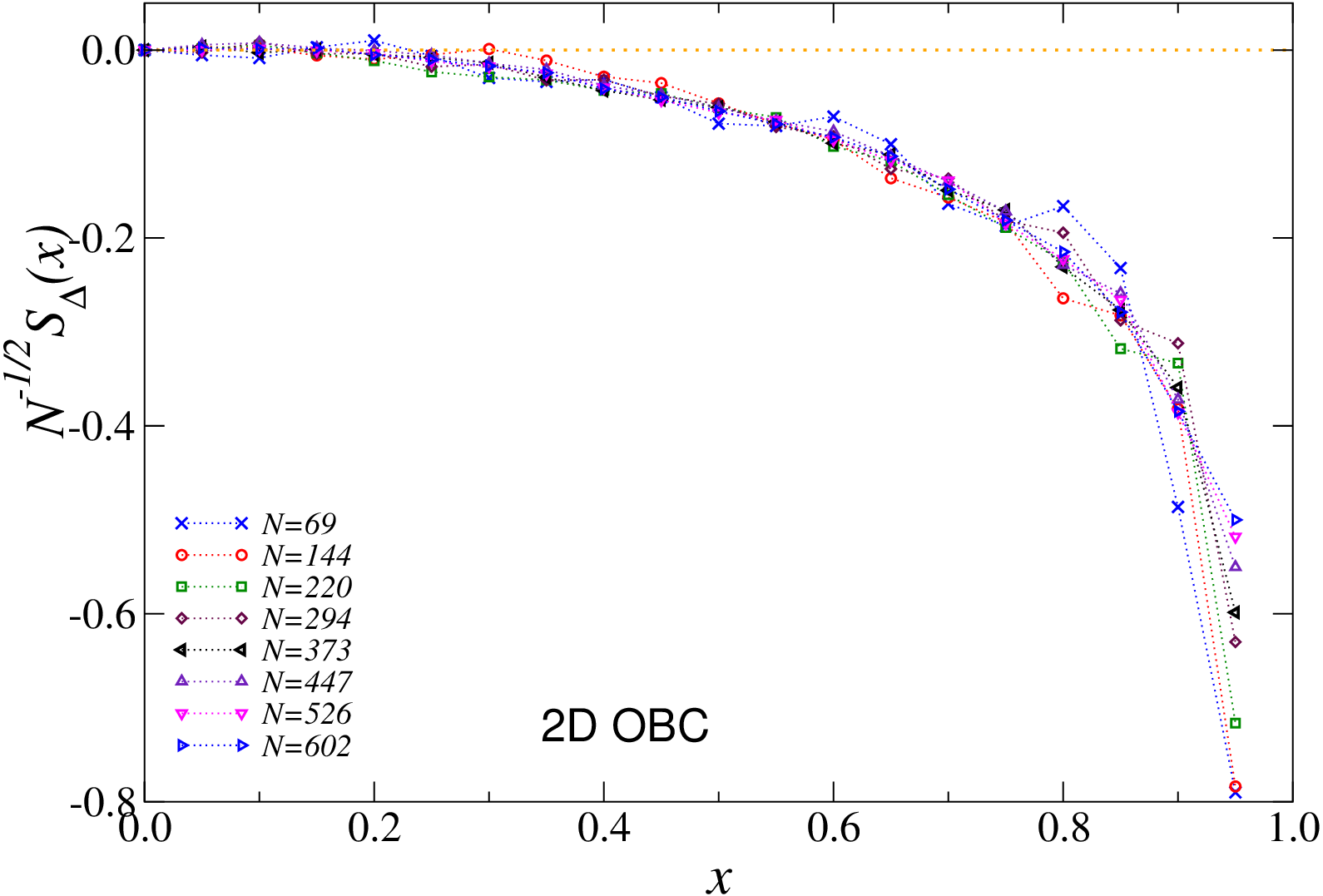}
\includegraphics*[scale=\graphicscale]{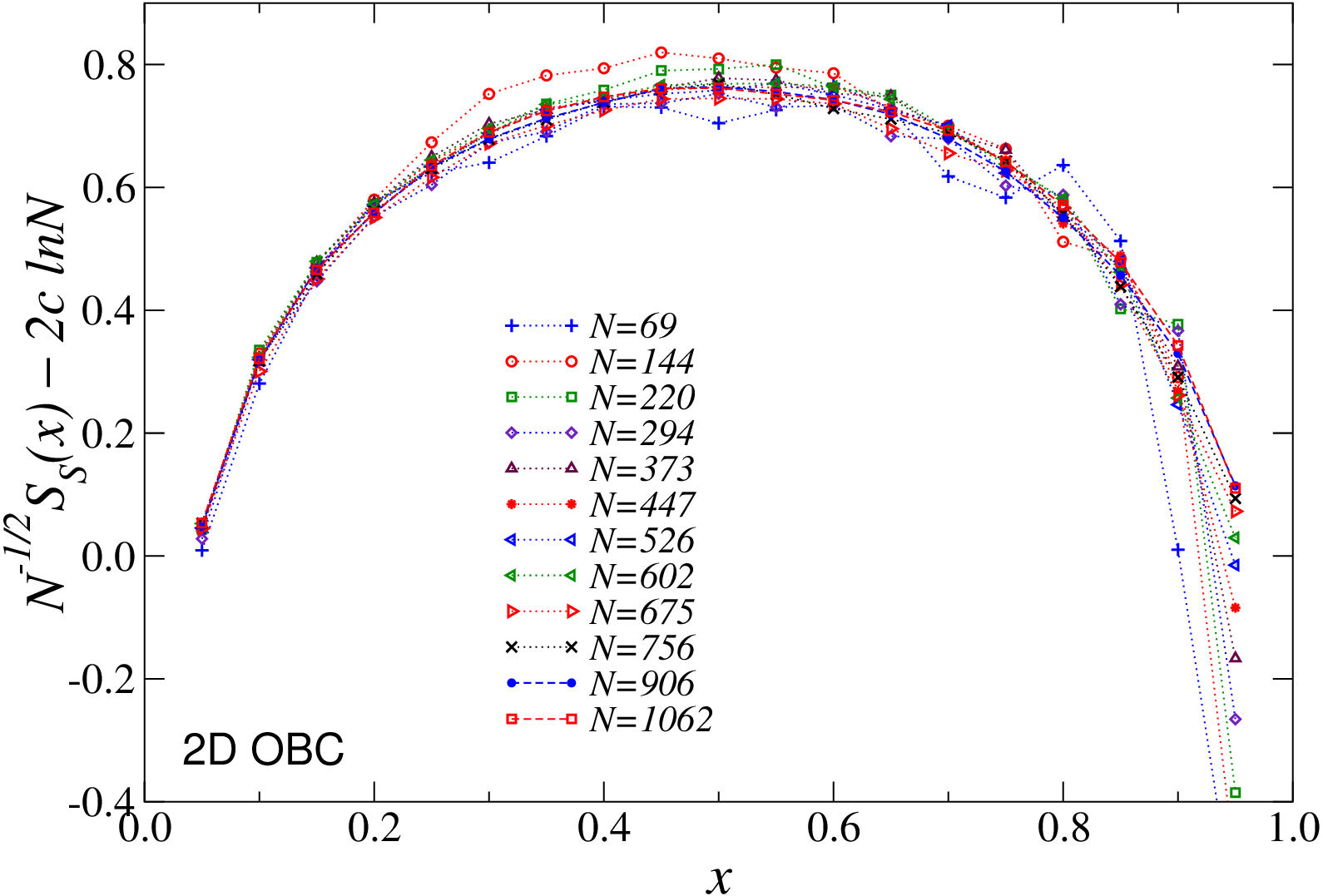}
\caption{(Color online) Bipartite vN entanglement entropies
$S_\Delta(x)$ (top) and $S_S(x)$ (bottom) for OBC as defined in
Eqs. (\ref{uxobca}) and (\ref{wobca}).
}
\label{uwplot}
\end{figure}

{\it Open boundary conditions}.  The results for the asymptotic
large-$N$ behavior of the entanglement entropy can be extended to OBC.
For simplicity, we consider only 2D systems.  We first study the
half-space entanglement entropy, i.e. when $A$ corresponds to the
region $z_1<L/2$.  The Widom conjecture gives
\begin{eqnarray}
S_{\rm HS}^{(\alpha)}  = N^{1/2}\left[ {1+\alpha^{-1}\over 12\pi^{1/2}} 
\ln N + O(1)\right],\label{uobca}
\end{eqnarray}
where the coefficient of the leading term is half of the one for PBC,
due to the fact that the boundary between the half spaces is half of
the one of stripes.  Fig.~\ref{ee2dobhs} show results at fixed $N$ up
to $N\approx 10^3$, for the half-space entanglement entropies $S_{\rm
HS}^{(\alpha)}$ showing agreement with the prediction and the
presence of sizable oscillating corrections as for PBC.

The presence of boundaries breaks translational invariance. Thus the
entanglement entropy depends on the location of the subsystem
$A$.  We consider the case when $A$ corresponds to the region
limited by an hyperplane parallel to one axis and at distance
$x+L/2$ from the boundary. We denote the corresponding
entanglement entropy by $S_B^{(\alpha)}(x)$ (with
$S_{{\rm HS}}^{(\alpha)} = S_{B}^{(\alpha)}(0)$).  Fig.  \ref{uwplot}
reports the quantity
\begin{equation}
 S_{\Delta}^{(\alpha)}(x)  \equiv  S_{B}^{(\alpha)}(x) - S_{B}^{(\alpha)}(0)  =  
 N^{1/2} \left[ f_\Delta^{(\alpha)}(x) + o(1) \right],
 \label{uxobca}
\end{equation}
where the function $f_{\Delta}(x)$ is a $O(1)$ contribution and cannot
be obtained from Widom conjecture.  Fig.  \ref{uwplot} shows clearly
that, despite of the oscillating corrections, the data for many
different $N$ collapse on the same master curve supporting the scaling
ansatz.

We also consider the case of a subsystem $A$ enclosed by two parallel
hyperplanes at the same distance $x<L/2$ from the center, 
both parallel to one of the cubic axes.  The corresponding entanglement
entropies are denoted by $S_{S}^{(\alpha)}(x)$.  Using again the Widom
conjecture for the leading large-$N$ behavior, we expect
\begin{equation}
 S_S^{(\alpha)}(x) =  N^{1/2}\left[ {1+\alpha^{-1}\over 6\pi^{1/2}}  
\ln N + f_S^{(\alpha)}(x) + o(1)\right],
\label{wobca}
\end{equation}
where the coefficient of the leading term equals the one for PBC.
Again the functions $f_S(x)$ cannot be obtained from Widom conjecture,
but the validity of this ansatz is supported by the data reported in
Fig.  \ref{uwplot}.

{\it Conclusions}.  We have studied the vN and R\'enyi entanglement
entropies of a system of $N$ particles in a volume $L^d$ with periodic
and open boundary conditions.  The large $N$ asymptotics of these
entanglement entropies can be obtained by means of 
Widom conjecture~\cite{Widom-81}
and can be written in the simple form
\begin{equation}
S^{(\alpha)}= \frac{1+\alpha^{-1}}{2} \frac{c_S}2 {\cal A}_z N^{1-1/d}\ln N
\label{Saconcl}
\end{equation}
where the coefficient $c_S$ is given in Eq. (\ref{resca2d2}) while
${\cal A}_z$ is the area between $A$ and $B$ in units where $L=1$.

This large-$N$ asymptotic behaviors can be also turned into an
asymptotic size dependence of the entanglement entropies for connected
bipartitions of fermion gases in a cubic box of volume $L^d$, in the
thermodynamic limit when $N,L\to\infty$ and $N/L^d=\rho$ with $\rho$
the particle density.  Eq. (\ref{Saconcl}) can be written as
\begin{eqnarray}
\frac{4 S^{(\alpha)}}{1+\alpha^{-1}}
\approx  {\cal A}_z  c_S \rho^{1-1/d} L^{d-1} \ln L\approx
  \frac{c_S}{d-1} \rho^{1-1/d}  {\cal A} \ln {\cal A}\,, \nonumber 
\end{eqnarray}
where we used $L^{d-1} {\cal A}_z={\cal A}$ with ${\cal A}$ is the
area separating $A$ and $B$.  This confirms the presence of general
logarithmic corrections to the area law in free fermion gases.

While the leading asymptotic behavior in $N$ of $S^{(\alpha)}$ is
understood from our analysis, the numerical data show the presence of
oscillating subleading corrections to the scaling whose structure and
origin is still not clear.

{\it Acknowledgements}. 
PC research was supported by ERC under the Starting Grant  n. 279391 EDEQS.
He also acknowledges the kind hospitality at the Institute Henri Poincar\'e, Paris.

\end{document}